# Detecting Algorithmically Generated Domains Using a GCNN-LSTM Hybrid Neural Network

Zheng Wang [1]

**Abstract**: Domain generation algorithm (DGA) is used by botnets to build a stealthy command and control (C&C) communication channel between the C&C server and the bots. A DGA can periodically produce a large number of pseudo-random algorithmically generated domains (AGDs). AGD detection algorithms provide a lightweight, promising solution in response to the existing DGA techniques. In this paper, a GCNN (gated convolutional neural network)-LSTM (long short-term memory) Hybrid Neural Network (GLHNN) for AGD detection is proposed. In GLHNN, GCNN is applied to extract the informative features from domain names on top of LSTM which further processes the feature sequence. GLHNN is experimentally validated using representative AGDs covering six classes of DGAs. GLHNN is compared with the state-of-the-art detection models and demonstrates the best overall detection performance among these tested models.

## 1. Introduction

Domain Name System (DNS) is a critical component of the Internet, providing indispensable naming and addressing support for billions of hosts and users. At the same time, DNS abuse fuels cyber-attacks that leverage the DNS to evade countermeasures. A notable example is Domain Generating Algorithms (DGAs) associated with malware like botnets [1], [2]. A widespread botnet in recent years was observed to use algorithmically generated domains (AGDs) as rendezvous points where the infected hosts and the command and control (C&C) servers communicate [3]. Botnets are often employed as a substrate to launch large-scale malicious activities such as executing DDoS attacks, sending spam, exfiltrating data, and conducting phishing campaigns. e.g., based on the telemetry data from a security company, a DDoS attack in 2020 [4] was coordinated through a botnet with 402,000 different IPs and a peak traffic low of 292,000 requests per second.

Malicious actors may take advantage of a specific domain name to let the compromised hosts locate the C&C server by looking up the domain name. However, if defenders identify the fixed domain name, they can quickly block the malware that depends on that domain name. In response, attackers use two tactics to evade detection and takedown of their malware. The first tactic, known as domain fluxing, is dynamically switching the domain names of the C&C server. If the domain name is changing at a smaller time interval, it will generally have a better chance of outpacing the detection by defenders. So defenders may find it difficult to block the rapidly switching C&C domain names. It is often likely that even if a C&C domain name is identified by defenders, it is already not in use and changed to a new domain. The second tactic is to create an asymmetry between attacker and defender. That is, a large number of candidate domain names are generated for malicious activities while only a few of them are actually used. The asymmetry makes it challenging to remediate the malware because it is much less costly to generate domain names by attackers than to track and ascertain malicious domain names by defenders. Both the two evasion tactics call for DGAs which use an algorithm and seed to automatically generate a pseudo-random sequence of characters to form a domain name. The DGA algorithm and the seed are shared beforehand between the attacker and the compromised hosts. Consequently, the AGD generation is synchronized between both sides without the need for communication. This allows the attacker to foreknow the domain name sequence that the compromised hosts will query for. Then the attacker can register one domain name from the sequence as the C&C domain name in advance. To locate the C&C server, the compromised host issues the DNS requests following the predetermined AGD sequence. It may persistently receive a domain-nonexistence response to each AGD until the C&C domain name is queried and successfully resolved to the IP addresses of the C&C server. In this way, the C&C communication channel is established between the compromised host and the C&C server and they can communicate directly since then. As a typical combination of domain fluxing and asymmetry, the attacker can periodically execute a DGA algorithm to generate a list of domain names and select one from the list as the new C&C rendezvous point.

[1] Email: zhengwang98@gmail.com

Detecting AGDs is among the best practices for detecting and protecting against DGA-fueled malware. Once the AGDs are correctly identified from the DNS traffic, they can be controlled, blocked, blacklisted, or taken down. Without the valid AGDs as the rendezvous point, the malware will lose the C&C communication channel and thereby get neutralized. In theory, malicious AGDs should be structured more or less differently from legitimate domains. e.g., some malicious domains may demonstrate seemingly random strings of characters whereas some legitimate domains often show more similarity to natural words/phases.

In the last decade, security researchers performed statistical analysis or used machine learning and artificial intelligence to identify AGDs from non-malicious domains. e.g,, some statistical features of domain name strings [5] and deep learning models [6]-[9] were proposed to detect AGDs. This work aims at advancing the AGD detection capability by proposing a new deep learning model. Specifically, a GCNN (gated convolutional neural network [10])-LSTM (long short-term memory [11]) Hybrid Neural Network (GLHNN) for AGD detection is proposed. The proposed GLHNN is evaluated in comparison with FANCI [5](the representative conventional machine learning-based detection approach) and five state-of-the-art deep learning-based detection models [6]-[9], [12]. Eleven representative DGAs of different classes in the DGA taxonomy are used in the evaluations. The results show that GLHNN outperforms the existing detection models in terms of overall detection performance.

The rest of the paper is organized as follows. The neural network architecture of GLHNN is presented in Section 2. We present the evaluation in Section 3. In Section 4, the related work on AGD detection are analyzed. Finally, Section 5 concludes the paper.

## 2. Neural Network Architecture

In this section, we present GLHNN in detail. We first outline the design, then detail the neural network layers, and finally present the data flow mathematically in the forward propagation direction.

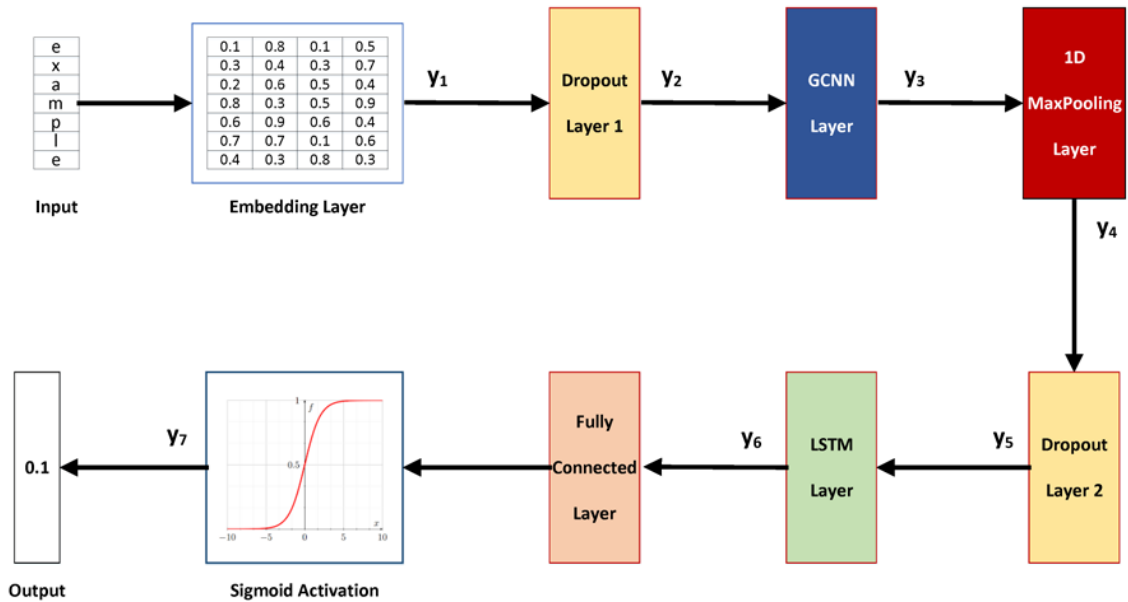

Fig. 1 GLHNN Architecture

## 2.1 Design

Fig. 1 shows the GLHNN Architecture. We first transform the input sequence to the embedding space through an embedding layer. The embedding provides semantic meaningful representation which is helpful to AGD prediction. As the core design of sequence feature extraction, we use a GCNN layer followed by a max-pooling layer on top of an LSTM layer. Similar to CNN, GCNN can merge information from adjacent tokens over the sequence. The advantage of GCNN over CNN is the gating mechanism. The gating mechanism allows the model to learn the relevance of features for predicting an AGD and accordingly control the information propagated to the next layer. Combined with a max-pooling layer for downsampling, GCNN serves to capture local sequence patterns effectively. With the local features as input, an LSTM layer accounts for sequential dependencies and extracts global features. The output of the LSTM layer is processed by a fully connected layer that captures the relationships between the different features. Finally, the fully connected layer is connected to a sigmoid activation which transforms the output into an informative prediction probability.

## 2.2 Neural Network Layers

*Embedding Layer*
An embedding is a machine-trainable mapping of a discrete categorical input variable to a vector of continuous numbers [13]. If the input to the neural network has the categorical features (e.g., words from a closed vocabulary of | vocab | distinct words), we need to encode such categorical features for use by the neural network. Generally, the encoding means associating each possible feature value (i.e., each word in the vocabulary) with a $d_{emb}$-dimensional vector. The encoding can be parameterized as a matrix $\mathbf{E} \in \mathbb{R}^{|vocab| \times d_{emb}}$ where each row corresponds to a different feature in the vocabulary. One-hot encoding assigns a unique dimension for each possible feature value and the resulting feature vector has a vector of 0s and a single 1 signaling the specific feature. The dimensionality of one-hot vector is the same as the number of distinct features such that $d_{emb} =$ | vocab | and the encoding matrix becomes a constant identity matrix $\mathbf{E} = \mathbf{I}_{|vocab|}$. Consequently, one-hot encoding suffers from high-cardinality feature issues because neural networks commonly do not manage very high-dimensional, sparse vectors well. Another drawback of one-hot encoding is the lack of semantic meaning. One-hot representations of features are completely independent of one another since the distances between one another are equal. For example, the word "cat" and "dog", despite some semantic similarity in a corpus, is not closer to each other than the word "cat" and "book" by one-hot encoding. Embedding overcomes the limitations of one-hot encoding by using dense and low-dimensional encoding vectors and make the encoding matrix trainable in a supervised-learning task. The encoding matrix $\mathbf{E}$ whose $d_{emb} <$ | vocab | is considered a parameter of the neural network and is trained jointly with the other parameters by supervision. The resulting embedded vectors are representations of features in an embedding space where similar features relative to the task are closer to one another. For example, the word "cat" may have a neighbor word "dog" in an embedding space because their semantic similarity can be captured by a supervised-learning task (using similar context, co-occurrence, etc.). The trainable semantic-relevant representation and a reduced encoding dimensionality, which are both offered by embedding, improve the generalization performance of the neural network.

Let the input sequence $\overline{\mathbf{x}}$ have $L$ tokens: $\overline{\mathbf{x}} = [\mathbf{x}_1; \dots; \mathbf{x}_L]^\top$ where each token $\mathbf{x}_i$ is encoded as a one-hot vector. Let $e_i$ denote the vector with a 1 in the $i$ th coordinate and 0's elsewhere, e.g., in $\mathbb{R}^5$, $e_3 = (0,0,1,0,0)$. We can re-sort the vocabulary to ensure the one-hot encoded tokens can be denoted by $\mathbf{x}_i = e_i^\top$. Then the embedding operation can be implemented as

$$\mathbf{y} = \overline{\mathbf{x}} \mathbf{E}$$

$$\mathbf{y} \in \mathbb{R}^{L \times d_{\text{emb}}}, \; \bar{\mathbf{x}} \in \mathbb{R}^{L \times |\text{vocab}|}, \; \mathbf{E} \in \mathbb{R}^{|\text{vocab}| \times d_{\text{emb}}} \tag{1}$$

*Dropout Layer*

Neural networks are prone to overfitting. Dropout is a technique used to improve overfitting on neural networks by randomly setting neurons to 0 with a rate during training. The Dropout layer can be implemented as a mask that nullifies the contribution of some neurons towards the next layer and leaves unmodified all others. The output vector of a Dropout layer **y** can be expressed as:

$$\begin{aligned} \mathbf{y} &= \mathbf{m} \odot \mathbf{x} \\ \mathbf{m} &\sim \text{Bernouli}(p) \end{aligned} \tag{2}$$

where **m** is a random masking vector with the dimension of the input vector **x** and the operator $\odot$ is an elementwise product. The values of the elements in **m**, which are either 0 or 1, are determined by a Bernouli distribution with parameter $p$. The values in the input vector corresponding to zeros in the masking vector are then nullified so that their incoming and outgoing connections in the neural network are also removed from the training. Note that the Dropout layer only applies when training and is deactivated during inference.

*1D Max-Pooling Layer*

Consider a sequence of tokens $x_{1:n} = x_1, \dots, x_n$, each with their corresponding $d_{\text{emb}}$ dimensional token embedding $\mathbf{E}_{[x_i]} = \mathbf{x}_i = [b_{1,i}, \dots, b_{d_{\text{emb}},i}]^\top$. A 1D max-pooling of size-$k$ works by moving a sliding-window of size $k$ over the sequence, and taking the maximum over each window in the sequence. Define the operator $\oplus (x_{i:i+k-1})$ to be the concatenation of the vectors $x_i, \dots, x_{i+k-1}$. The concatenated vector of the $i$ th window is then $\tilde{\mathbf{x}}_i = \oplus (\mathbf{x}_{i:i+k-1}) = [\mathbf{x}_i; \dots; \mathbf{x}_{i+k-1}], \tilde{\mathbf{x}}_i \in \mathbb{R}^{k \times d_{\text{emb}}}$. The max-pooling over one pooling window can be written as:

$$\mathbf{y}_i = \mathbf{M}^{Pool}(\tilde{\mathbf{x}}_i) = [g_{i,1}, \dots, g_{i,d_{\text{emb}}}] \tag{3}$$

$$g_{i,j} = \max_{i \leq m \leq i+k-1} b_{j,m} \quad \forall j \in [1, d_{\text{emb}}] \tag{4}$$

$$\mathbf{y}_i \in \mathbb{R}^{d_{\text{emb}}}, \; \tilde{\mathbf{x}}_i \in \mathbb{R}^{k \times d_{\text{emb}}}$$

where $g_{i,j}$ denotes the $j$ th component of $\mathbf{y}_i$. By moving the pooling windows with a stride, we can obtain a number of pooling outputs. By stacking these outputs over all pooling windows, the final output is represented as:

$$\mathbf{y} = \mathcal{M}^{Pool}(\bar{\mathbf{x}}) = [\mathbf{y}_1^\top; \dots; \mathbf{y}_S^\top]^\top \tag{5}$$

$$\mathbf{y} \in \mathbb{R}^{S \times d_{\text{emb}}}, \; \bar{\mathbf{x}} \in \mathbb{R}^{L \times d_{\text{emb}}}$$

where the input sequence $\bar{\mathbf{x}}$ has $L$ tokens: $\bar{\mathbf{x}} = [\mathbf{x}_1; \dots; \mathbf{x}_L]^\top$ and $S$ is the number of sliding windows. $S$ is jointly determined by the pooling size $k$ and the pooling stride $r$:

$$S = (L - k + 1)/r \tag{6}$$

Some tokens on both sides of the input sequence may not be covered by the sliding pooling windows. To avoid losing the information contained in the border tokens, padding is added to the sides of the input sequence to allow more length for the pooling windows to cover the sequence. One padding strategy for max-pooling is to pad evenly to the left/right of the input sequence with a value of $-inf$ to ensure the output has the same dimension as the input if a stride of 1 is applied. For that "same"padding strategy, the number of padding tokens is $k-1$ per Eq.(6) so that Eq.(6) is turned into:

$$S_{same} = L/r \tag{7}$$

*GCNN Layer*
Consider the input sequence $\bar{\mathbf{x}}$ has $L$ tokens: $\bar{\mathbf{x}} = [\mathbf{x}_1; ...; \mathbf{x}_L]^T$ where each $\mathbf{x}_i$ has its corresponding $d_{emb}$ dimensional token embedding $\mathbf{E}_{[x_i]} = \mathbf{x}_i = [b_{1,i}, ..., b_{d_{emb},i}]^T$. The 1D convolution operation applies a set of kernel functions $\mathbf{C}^{(l)} = \left[c_{i,j}^{(l)}\right] \in \mathbb{R}^{k \times d_{emb}}$ over the 1D windows of $k$ size which slide along the input sequence with a stride $r$. Let the number of kernel functions be $d_{out}$ and we have the kernel matrix $\mathbf{C} = \left[\mathbf{C}^{(1)}; ...; \mathbf{C}^{(d_{out})}\right]$. 1D convolution is denoted by the operator $*$ and is defined as:

$$\mathbf{y}^c = \bar{\mathbf{x}} * \mathbf{C} \quad \Rightarrow \quad y_{m,l} = \sum_{n=1}^{k} \sum_{d=1}^{d_{emb}} b_{m-k+n,d} \times c_{n,d}^{(l)}$$

$$\mathbf{y}^c = [y_{i,j}] \in \mathbb{R}^{S \times d_{out}}, \quad \bar{\mathbf{x}} \in \mathbb{R}^{L \times d_{emb}}, \quad \mathbf{C} \in \mathbb{R}^{k \times d_{emb} \times d_{out}} \tag{8}$$

Here at each position $m$, $1 \leq m \leq L$, the elementwise product $\mathbf{C}^{(k)} \odot [\mathbf{x}_{m-k+1}; ...; \mathbf{x}_m]$ is computed and then summed. Let the convolution adopt a sliding window covering the previous $k-1$ tokens and the current token at each position $m$ so that the kernels are prevented from seeing future tokens. Accordingly, we zero-pad the input sequence with $k-1$ tokens preceding the first token to ensure the output has the same length as the input when the stride is 1. Here we set the stride $r = 1$. So we have the number of the sliding windows $S = L$ (see Eq.(7)).

As illustrated in Fig. 2, a GCNN layer is composed of a convolutional block that produces two separate convolutional outputs ($G_1$ and $G_2$ in Fig. [2]), and a gating operation block that uses one convolutional output ($G_1$) to gate the other ($G_2$) [10]. A residual skip connection is also added to straightforwardly direct the input to the output. To align the dimensions of the output with those of the input, let the number of the kernels match the embedding dimension: $d_{out} = d_{emb}$. The output of a GCNN layer is represented as:

$$\mathbf{y} = \bar{\mathbf{x}} + (\bar{\mathbf{x}} * \mathbf{W} + \mathbf{b}) \otimes \text{sigm}(\bar{\mathbf{x}} * \mathbf{V} + \mathbf{d})$$

$$\mathbf{y}, \bar{\mathbf{x}}, \mathbf{b}, \mathbf{d} \in \mathbb{R}^{L \times d_{emb}}, \quad \mathbf{W}, \mathbf{V} \in \mathbb{R}^{k \times d_{emb} \times d_{emb}} \tag{9}$$

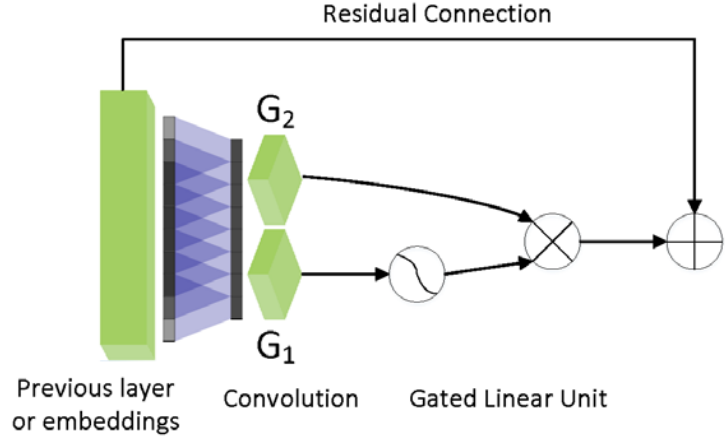

Fig. 2 GCNN Layer

where $\mathbf{W}, \mathbf{V}$ are the kernel matrices and $\mathbf{b}, \mathbf{d}$ are the biases. Here the gating mechanism is important because it allows selection of the tokens or features that are important for predicting the AGD, and provides a mechanism to learn and pass along just the relevant information. Like activation functions, the gating mechanism also provides the layer with non-linear capabilities. By providing a highway for the gradient during backpropagation, the residual connection minimizes the vanishing gradient problem, allowing networks to be built much deeper.

### *LSTM Layer*
The long short-term memory (LSTM) [11] is a popular variant of recurrent neural networks (RNNs) that is more robust to the gradient vanishing and gradient exploding problems. The LSTM state after reading token $t$ is composed of two vectors $(\mathbf{h}_t, \mathbf{c}_t)$ where $\mathbf{h}_t$ is the hidden state vector and $\mathbf{c}_t$ is the cell state vector. The recurrent operation of LSTM can be represented as:

$$(\mathbf{h}_t, \mathbf{c}_t) = \mathbf{F}^{LSTM}\big(\mathbf{x}_t, (\mathbf{h}_{t-1}, \mathbf{c}_{t-1})\big) \qquad [10]$$

which is mathematically defined as:

$$\begin{bmatrix}\mathbf{i}_t\\ \mathbf{f}_t\\ \mathbf{o}_t\\ \tilde{\mathbf{c}}_t\end{bmatrix} = \begin{pmatrix}\text{sigm}\\ \text{sigm}\\ \text{sigm}\\ \tanh\end{pmatrix} \quad \mathbf{W}^{LSTM}\begin{bmatrix}\mathbf{x}_t\\ \mathbf{h}_{t-1}\end{bmatrix} \qquad [11]$$

$$\mathbf{c}_t = \mathbf{f}_t \odot \mathbf{c}_{t-1} + \mathbf{i}_t \odot \tilde{\mathbf{c}}_t \qquad [12]$$

$$\mathbf{h}_t = \mathbf{o}_t \odot \tanh(\mathbf{c}_t) \qquad [13]$$

where the update matrix is denoted by $\mathbf{W}^{LSTM}$ and the four intermediate vector variables are functions of the input and previous hidden state in Eq.(12). Three of them, namely $\mathbf{i}_t, \mathbf{f}_t$, and $\mathbf{o}_t$, are known as the input, forget, and output gates respectively. These gates are computed from sigmoid activations, $\text{sigm}(x) = \frac{1}{e^{-x}+1}$, ensuring that their values will be in the range [0,1]. An update candidate $\tilde{\mathbf{c}}_t$ can be considered as a representation of the input transformed by a tanh activation function, $\tanh(x) = \frac{e^{2x}-1}{e^{2x}+1}$. The operator $\odot$ is an elementwise product. When the cell state is updated in Eq.([cell]), the forget gate $\mathbf{f}_t$ controls how much of the current cell state should be forgotten and the input gate $\mathbf{i}_t$ controls how much of the new input (contained in $\tilde{\mathbf{c}}_t$) should be written to the cell memory. Finally, the hidden state is updated in Eq.(13) where the cell state is passed through a tanh activation function and controlled by the output gate $\mathbf{o}_t$. The design of cell memory and gating mechanisms allow gradient information to propagate through the network over long distances of units.

The hidden state is taken as the output:

$$\mathbf{y}_t = \mathbf{h}_t \quad [14]$$

Then following Eq.(11) recursively, we have the output of the last LSTM unit as the final output of LSTM:

$$\mathbf{y} = \mathbf{y}_L = \mathcal{F}_h^{LSTM}\big(\overline{\mathbf{x}}, (\mathbf{h}_0, \mathbf{c}_0)\big)$$

$$\mathbf{y} \in \mathbb{R}^{d_{emb}}, \overline{\mathbf{x}} \in \mathbb{R}^{L \times d_{emb}} \quad [15]$$

where the input sequence $\overline{\mathbf{x}}$ has $L$ tokens: $\overline{\mathbf{x}} = [\mathbf{x}_1; \ldots; \mathbf{x}_L]^\top$ and the final output is a function of the input sequence and the initial states $(\mathbf{h}_0, \mathbf{c}_0)$.

### *Fully Connected Layer and Activation Function*

A fully connected layer connects every neuron in the input to every neuron in the output. So the output is simply the linear transformation of the input:

$$\mathbf{y} = \mathbf{x}\mathbf{W} + \mathbf{b}$$
$$\mathbf{x} \in \mathbb{R}^{d_{in}}, \quad \mathbf{W} \in \mathbb{R}^{d_{in} \times d_{out}}, \quad \mathbf{b} \in \mathbb{R}^{d_{out}}, \quad \mathbf{y} \in \mathbb{R}^{d_{out}} \quad [16]$$

where $\mathbf{W}$ is the weight matrix and $\mathbf{b}$ is the bias.

A linear transformation is limited in modeling complex functions that do not follow linearity straightforwardly. So we need to introduce non-linearity to the neural networks. By attaching a non-linear activation function to the output of fully connected networks, the neural network ability to approximate a non-linear model is improved. When a non-linear element-wise activation function $\mathcal{G}$ is applied, we have Eq.(16) updated as:

$$\mathbf{y}^{Act} = \mathcal{G}(\mathbf{x}\mathbf{W} + \mathbf{b})$$
$$\mathbf{x} \in \mathbb{R}^{d_{in}}, \quad \mathbf{W} \in \mathbb{R}^{d_{in} \times d_{out}}, \quad \mathbf{b} \in \mathbb{R}^{d_{out}}, \quad \mathbf{y}^{Act} \in \mathbb{R}^{d_{out}} \quad [17]$$

Here we use the sigmoid activation for $\mathcal{G}$:

$$\mathcal{G}(x) = \text{sigm}(x) = \frac{1}{e^{-x}+1} \quad [18]$$

Since we need to predict the AGD probability as an output, the output of the neural network should fall into the range [0,1]. Sigmoid can serve the task well by converting the fully connected layer's output into a probability score.

## 2.3 Data Flow

Fig. 1 illustrates the input and output of each layer. Let the input sequence $\bar{\mathbf{x}}$ have $L$ tokens: $\bar{\mathbf{x}} = [\mathbf{x}_1; \ldots; \mathbf{x}_L]^\top$ where each token $\mathbf{x}_i$ is encoded as a one-hot vector. Then the output of Embedding Layer is (see Eq.(1)):

$$\mathbf{y}_1 = \bar{\mathbf{x}} \mathbf{E}$$

$$\mathbf{y}_1 \in \mathbb{R}^{L \times d_{emb}}, \quad \bar{\mathbf{x}} \in \mathbb{R}^{L \times |vocab|}, \quad \mathbf{E} \in \mathbb{R}^{|vocab| \times d_{emb}} \quad [19]$$

Then the output vector of Dropout Layer 1 can be expressed as (see Eq.(2)):

$$\begin{aligned} \mathbf{y}_2 &= \mathbf{m}_1 \odot \mathbf{y}_1 \\ \mathbf{m}_1 &\sim \text{Bernouli}(p_1) \end{aligned}$$

$$\mathbf{y}_1, \mathbf{y}_2, \mathbf{m}_1 \in \mathbb{R}^{L \times d_{emb}} \quad [20]$$

Next, the output of GCNN Layer is (see Eq.(9)):

$$\mathbf{y}_3 = \mathbf{y}_2 + (\mathbf{y}_2 * \mathbf{W} + \mathbf{b}) \otimes \text{sigm}(\mathbf{y}_2 * \mathbf{V} + \mathbf{d})$$

$$\mathbf{y}_3, \mathbf{y}_2, \mathbf{b}, \mathbf{d} \in \mathbb{R}^{L \times d_{emb}}, \quad \mathbf{W}, \mathbf{V} \in \mathbb{R}^{k \times d_{emb} \times d_{emb}} \quad [21]$$

Then the output of 1D Max-Pooling Layer is (see Eq.(5)):

$$\mathbf{y}_4 = \mathcal{M}^{Pool}(\mathbf{y}_3)$$

$$\mathbf{y}_4 \in \mathbb{R}^{S \times d_{emb}}, \quad \mathbf{y}_3 \in \mathbb{R}^{L \times d_{emb}} \quad [22]$$

where $S = L/r$ with the "same" padding (see Eq.(7)).

Then the output vector of Dropout Layer 2 can be expressed as (see Eq.(2)):

$$\begin{aligned} \mathbf{y}_5 &= \mathbf{m}_2 \odot \mathbf{y}_4 \\ \mathbf{m}_2 &\sim \text{Bernouli}(p_2) \end{aligned}$$

$$\mathbf{y}_5, \mathbf{y}_4, \mathbf{m}_2 \in \mathbb{R}^{S \times d_{emb}} \quad [23]$$

Next, the output vector of LSTM Layer is (see Eq.(15)):

$$\mathbf{y}_6 = \mathcal{F}_h^{LSTM}(\mathbf{y}_5, (\mathbf{h}_0, \mathbf{c}_0))$$

$$\mathbf{y}_6 \in \mathbb{R}^{d_{emb}}, \quad \mathbf{y}_5 \in \mathbb{R}^{S \times d_{emb}} \quad [24]$$

Finally, the output of Fully Connected Layer and then Sigmoid Activation is (see Eq.(17) and Eq.(18)):

$$\mathbf{y}_7 = \text{sigm}(\mathbf{y}_6 \mathbf{W} + \mathbf{b})$$

$$\mathbf{y}_6 \in \mathbb{R}^{d_{emb}}, \quad \mathbf{W} \in \mathbb{R}^{d_{emb} \times 1}, \quad \mathbf{b} \in \mathbb{R}^1, \quad \mathbf{y}_7 \in \mathbb{R}^1 \quad [25]$$

# 3. Evaluation

This section presents the evaluation of the proposed GLHNN. The domain name datasets used in the evaluation are described. A DGA taxonomy is used to select some representative DGAs from each DGA class. A brief description of the AGD detection algorithms used in comparison with GLHNN is presented. The experiment setting of the evaluation and the evaluation metrics are introduced. The results of the evaluation are presented and some findings are discussed.

## 3.1 Datasets

For the binary classification task, we need both a benign domain name dataset and an AGD dataset. The benign domain name dataset represents the non-malicious domain names observed on the Internet. The AGD dataset includes the malicious domain names generated by different DGAs.

***Benign Domain Name Dataset***. The most popular 1 million domain names ranked by Alexa [15] are used as benign domain names in the evaluation. The Alexa rank is calculated using a combination of estimated average daily unique visitors to the site and the estimated number of page views on the site over the past 3 months. A malicious site is unlikely to persistently survive detection and remediation. The domain names in the Alexa set are likely to attract the most web traffic and are considered benign.

***AGD Datasets***. In the previous study on DGA, some researchers obtained AGDs generated by some real-world DGAs. They were either collected on the Internet or produced by reverse engineering DGA code. Unlike those DGAs actually in use by malware, some new DGAs were also proposed in the cyber security literature. They were tested against the existing detection approaches to explore their capability and vulnerability. The evaluation uses both the real-world and human-crafted AGD sets to examine the universal relevance of GLHNN. Specifically, the dataset has the following sources:

- *DGArchive* is a free DGA repository offered by Fraunhofer FKIE and administrated by Daniel Plohmann. It is probably one of the most accessible and comprehensive AGD datasets, covering 98 DGAs and totaling some 117 million unique AGDs.

- *Hidden Markov Model (HMM)-based DGA* [16] is a human-crafted DGA proposed in academia. The researchers modeled the character sequence of a domain name string as an HMM process and used the legitimate domain name set to train the HMM model. A trained HMM model can generate new domain names which share the statistical properties of the training set. Consequently, the generated domain names may look like legitimate domain names, making them difficult to detect. In [16], the hyperparameters of the HMM model were tested to find the optimal in terms of anti-detection capability. The evaluation uses the optimal hyperparameters specified in  to obtain the AGD dataset.

- *DeepDGA* [17] is a deep learning-based human-crafted DGA. The deep learning algorithm first trains an auto-encoder model using the Alexa domain name dataset. Then the trained auto-encoder is split into an encoder and a decoder which are reassembled in a generative adversarial network. The generative adversarial network is used as a domain name generator. The deep learning architecture and the training methods are reproduced following [17]. The evaluation uses the AGDs dataset generated by the DeepDGA model.

## 3.2 DGA Taxonomy

A DGA taxonomy is used to classify all DGAs in the dataset. The features of the DGA taxonomy include seed source and generation scheme which were proposed in [18] as well as learning property:

**Seed Source**. As a pseudo--random generator, DGA computes its AGDs using a fixed determi-nistic algorithm and seeds actually determine a sequence of pseudo-random characters. That is, a DGA with the same seeds as parameters should produce the same AGD sequence. Seeding has two properties for a DGA:

- *Time dependence* means the DGA is time-based. Many DGAs attach a validity time to their AGDs. Others may incorporate time information into their seeds. The time dependence is often employed to make predicting future AGDs more difficult for defenders.

- *Determinism* characterizes the reproducibility of seeds. Some seedings use publicly accessible data such as the current time and the current exchange rates. This makes it possible for defenders to reproduce the AGDs in the aftermath of the attacks. Other seedings may use publicly inaccessible data (e.g., the exfiltrated personal information). Attackers can use non-deterministic seeding to protect from forensic DGA analysis in the aftermath of the attacks.

**Generation Scheme**. DGA uses seeds as input to its generation schemes. We have the following typical generation schemes:

- *Arithmetic-based DGAs* computes the AGD using an ASCII encoding or an offset sequence of a domain name.

- *Hash-based DGAs* generates the AGD by hashing the sequence of a domain name.

- *Wordlist-based DGAs* selects and concatenates words from a predetermined wordlist to form the AGD.

- *Permutation-based DGAs* develops permutations of a base domain name to generate the AGD.

**Learning Property**. Real-world DGAs are mostly determined by seeding and generation schemes. So they are generally not adaptive to the target domain name set that they are mimicking. The weakness is addressed by learning-based DGAs which can be adjustable by a training domain set. By fitting the DGA to the training set, the DGA can be self-tuned without human guidance.

***DGA Classes***. The DGA taxonomy is presented in Table 1. In Table 1, "XXY-Z-K" denotes a DGA class where "XXY", "Z", and "K" denotes seed source, generation scheme, and learning property respectively. Different combinations of the three features give different DGA classes. So in theory, there should be 32 DGA classes in total. However, only six DGA classes are identified from the DGArchive dataset plus the HMM-based DGA and DeepDGA, namely TDD-H-N, TDN-A-N, TDD-A-N, TDD-W-N, TID-A-N, and TID-A-L.

Table 1. DGA taxonomy

| DGA Class | Feature | Property | | Denotation |
|---|---|---|---|---|
| XXY-Z-K | Seed Source (XXY) | Time dependence (XX) | time-dependent | TD |
| | | | time-independent | TI |
| | | Determinism (Y) | deterministic | D |
| | | | non-deterministic | N |
| | Generation Scheme (Z) | Arithmetic-based | | A |
| | | Hash-based | | H |
| | | Wordlist-based | | W |
| | | Permutation-based | | P |
| | Learning Property (K) | Non-learning | | N |
| | | Learning | | L |

## 3.3 AGD Detection Algorithms

To validate GLHNN, the following detection algorithms are implemented and compared with GLHNN. The implementations use the architectures and hyperparameters in line with their respective literature.

***FANCI***. FANCI [5] is a proven conventional machine learning approach to detect malicious domain names. It is based on feature extraction from a domain name string. Then the features can be used by a classification algorithm to separate malicious domains from non-malicious ones. The effectiveness of FANCI highly depends on feature extraction.

***LSTM***. LSTM is a classic recurrent neural network that takes into consideration what it has learned from prior inputs to classify the current input. LSTM introduces additional signal information, namely cell memory, to the vanilla recurrent neural network. It uses the gate mechanism to cope with the vanishing gradient problem suffered by the vanilla recurrent neural network. LSTM is commonly used to process sequential data. In [6], an LSTM model was designed to classify domain names.

***Bidirectional LSTM (Bi-LSTM)***. Bi-LSTM is an extension of LSTM. LSTM only processes the sequence in one direction whereas Bi-LSTM incorporates the information in both directions. To achieve this, Bi-LSTM adds one more LSTM layer to standard LSTM, which reverses the information flow direction of the other LSTM layer. The bidirectional information flow can be combined to produce a more meaningful output. So Bi-LSTM has advantages over LSTM in many real-world problems. The Bi-LSTM model was proposed in [12] to address the AGD detection problem.

***CNN***. CNNs were initially designed for image processing tasks with well-recognized success [19], [20]. A CNN is typically comprised of two operations, which can be thought of as feature extractors: convolution and pooling. The convolution operation computes the output through the application of a kernel to the input data. The kernel slides with a defined stride through every element in the input data when an element-wise multiplication of the kernel and the input matrix is calculated. The resulting output is called a feature map which is smaller than the input but contains all important information. The pooling operation is responsible for detecting features irrespective of their location from a feature map. This is done by computing a certain operation of a feature map. Most common pooling includes max-pooling and average-pooling which computes the maximum and average value respectively within the sliding window. Ideally, the output of pooling will capture the most relevant features. The extracted features by convolution and pooling are typically fed to a fully connected layer that performs prediction. In the case of the AGD detection task, a domain name string is considered a 1-dimensional sequence. So CNN is also configured to do 1-dimensional convolution and pooling operations. The CNN implementation for AGD detection was proposed in [7].

***CNN-LSTM (hybrid CNN/LSTM-based neural networks)***. CNN-LSTM [8] is designed to combine the advantages of CNN and LSTM. It applies a CNN layer to the input as a feature extractor. The output feature map, which is also a sequential data, is then processed by an LSTM layer. The design is based on the understanding of the comparative capabilities and limitations of CNN and LSTM. Basically, CNN is more suitable to capture local information because it lacks the mechanism of recognizing the global information over long distances. By contrast, LSTM does better in extracting global information thanks to its memory mechanism.

***LSTM-Att (LSTM combined with an attention mechanism)***. Attention [21], [22] provides a mechanism to allow output selectively focus on certain elements of the sequential data. During the training, the extent of attention is learned through the attention weights assigned to each input element. And a context vector, which is the weighted sum of input elements, is used to make a prediction. Attention overcomes the limitation of LSTM in capturing long-term dependencies of a sequence. The LSTM-Att model for AGD detection was proposed in [9].

## 3.4 Experiment Setting

***Datasets***. Nine DGAs are selected to sample their AGDs from DGArchive [23]. Taking the HMM-based DGA [16] and DeepDGA [17] into account, we have eleven DGAs for the evaluation. Considering the DGA taxonomy in Section 3.2, the selected DGAs ensure each of the six DGA classes has one or two DGAs. They are classified as the following:

- *TDD-H-N*: *Bamital* [24], *Dyre* [25].

- *TID-A-N*: *Banjori* [26], *Ramdo* [27].
- *TDN-A-N*: *Bedep* [28].
- *TDD-A-N*: *Conficker* [29], *Pushdo* [30].
- *TDD-W-N*: *Matsnu* [31], *Suppobox* [32].
- *TID-A-L*: *HMM-based DGA* [16], *DeepDGA* [17].

**Preprocessing**. The AGD detection models cannot handle variable sequences. So longer sequences are truncated and shorter sequences are padded to ensure all preprocessed sequences have the same length. Here the fixed sequence length is set to 256. RFC 1035 [33] specifies the valid characters in a domain name, including the letters A-Z, the digits 0-9 and hyphen (-). So including the padding character, 39 unique characters can be contained in a domain name. Before being fed to the detection model, each domain name gets its Top Level Domain (TLD) removed. TLD represents the rightmost suffix of a domain name (e.g., "com" for the domain "example.com"), which can be considered irrelevant to the AGD detection task.

**Training**. We use random sample with replacement to generate the training and validation sets. The sampling is conducted 100 times to obtain 100 sets of 5000 samples for the *Alexa* set and each AGD set. Then for each AGD, we can have 100 pairs of training and validation sets, each of which contains one AGD set and one *Alexa* set. Stratified 5-fold cross-validation is performed on each pair of training and validation sets. So a total of 500 experiments are conducted for each DGA in the evaluation. The overall performance is obtained by averaging across the 500 experiments.

The detection algorithms used in the evaluation include the six detection algorithms described in Section 3.3 and GLHNN. Eleven DGAs are tested under each of the detection algorithms. Binary cross entropy is used as the loss function. The deep learning-based algorithms among the detection algorithms use the Adam algorithm [34] with a learning rate of 0.001 and a batch size of 256. The training is stopped when the accuracy is not improved over 10 epochs.

### 3.5 Evaluation Metrics

Binary classification metrics are used to evaluate the performance of AGD detectors. The definitions of the following four metrics are based on the confusion matrix in Table 2.

Accuracy (ACC) measures the ratio of correctly classified observations, both positive and negative:

$$ACC = \frac{TP+TN}{FP+FN+TP+TN} \qquad [26]$$

Table 2. Confusion matrix of classification

| Actual Class \ Predicted Class | Malicious | Benign |
|---|---|---|
| Malicious | True Positive (TP) | False Negative (FN) |
| Benign | False Positive (FP) | True Negative (TN) |

Precision (PRE) measures the ratio of correctly classified positive observations to the total classified positive observations:

$$PRE = \frac{TP}{TP+FP} \quad [27]$$

Recall (REC) measures the ratio of correctly classified positive observations to the total observations in the actual positive class:

$$REC = \frac{TP}{FN+TP} \quad [28]$$

F1 score is the weighted average of Precision and Recall:

$$F1 = 2 \times \frac{PRE \times REC}{PRE+REC} \quad [29]$$

## 3.6 Results

***Detecting Banjori-Generated AGDs***. Table 3 shows the AGD detection results for Banjori (which falls into the TID-A-N class). GLHNN demonstrates the best detection performance under the four evaluation metrics by accurately classifying all samples. And the six deep learning-based detection models outperform the conventional machine learning-based model, namely FANCI.

Table 3. AGD Detection Results For Banjori

| Model | Accuracy | Precision | Recall | F1 |
|---|---|---|---|---|
| FANCI | 0.9785 | 0.9787 | 0.9785 | 0.9785 |
| LSTM | 0.9940 | 0.9941 | 0.9940 | 0.9940 |
| Bi-LSTM | 0.9960 | 0.9960 | 0.9960 | 0.9960 |
| CNN | 0.9980 | 0.9980 | 0.9980 | 0.9980 |
| CNN-LSTM | 0.9980 | 0.9980 | 0.9980 | 0.9980 |
| LSTM-Att | 0.9890 | 0.9891 | 0.9890 | 0.9890 |
| GLHNN | **1.0000** | **1.0000** | **1.0000** | **1.0000** |

***Detecting Bedep-Generated AGDs***. Table 4 shows the AGD detection results for Bedep (which falls into the TDN-A-N class). GLHNN achieves the best detection performance under the four evaluation metrics. And the six deep learning-based detection models outperform FANCI.

Table 4. AGD Detection Results For Bedep

| Model | Accuracy | Precision | Recall | F1 |
|---|---|---|---|---|
| FANCI | 0.9465 | 0.9467 | 0.9465 | 0.9465 |
| LSTM | 0.9940 | 0.9941 | 0.9940 | 0.9940 |
| Bi-LSTM | 0.9800 | 0.9800 | 0.9800 | 0.9800 |
| CNN | 0.9885 | 0.9885 | 0.9885 | 0.9885 |
| CNN-LSTM | 0.9890 | 0.9890 | 0.9890 | 0.9890 |
| LSTM-Att | 0.9835 | 0.9835 | 0.9835 | 0.9835 |
| **GLHNN** | **0.9945** | **0.9945** | **0.9945** | **0.9945** |

***Detecting Conficker-Generated AGDs***. Table 5 shows the AGD detection results for Conficker (which falls into the TDD-A-N class). GLHNN achieves the best detection performance under the four evaluation metrics. And except for LSTM-Att, the deep learning-based detection models outperform FANCI.

Table 5. AGD Detection Results For Conficker

| Model | Accuracy | Precision | Recall | F1 |
|---|---|---|---|---|
| FANCI | 0.9105 | 0.9127 | 0.9105 | 0.9104 |
| LSTM | 0.9165 | 0.9176 | 0.9165 | 0.9164 |
| Bi-LSTM | 0.9385 | 0.9389 | 0.9385 | 0.9385 |
| CNN | 0.9525 | 0.9526 | 0.9525 | 0.9525 |
| CNN-LSTM | 0.9510 | 0.9510 | 0.9510 | 0.9510 |
| LSTM-Att | 0.8935 | 0.8990 | 0.8935 | 0.8931 |
| **GLHNN** | **0.9530** | **0.9533** | **0.9530** | **0.9530** |

***Detecting Matsnu-Generated AGDs***. Table 6 shows the AGD detection results for Matsnu (which falls into the TDD-W-N class). GLHNN is still the best under the four evaluation metrics. And the deep learning-based detection models outperform FANCI.

Table 6. AGD Detection Results For Matsnu

| Model | Accuracy | Precision | Recall | F1 |
|---|---|---|---|---|
| FANCI | 0.8560 | 0.8781 | 0.8560 | 0.8539 |
| LSTM | 0.8870 | 0.8880 | 0.8870 | 0.8869 |
| Bi-LSTM | 0.8870 | 0.8880 | 0.8870 | 0.8869 |
| CNN | 0.9035 | 0.9075 | 0.9035 | 0.9033 |
| CNN-LSTM | 0.9135 | 0.9148 | 0.9135 | 0.9134 |
| LSTM-Att | 0.8800 | 0.8814 | 0.8800 | 0.8799 |
| **GLHNN** | **0.9165** | **0.9211** | **0.9165** | **0.9163** |

***Detecting HMM-based-DGA-Generated AGDs***. Table 7 shows the AGD detection results for HMM-based DGA (which falls into the TID-A-L class). CNN-LSTM is the best model under the four evaluation metrics. GLHNN is less effective than the two deep learning-based models applying CNN, namely CNN and CNN-LSTM. The deep learning-based detection models outperform FANCI.

Table 7. AGD Detection Results For HMM-based DGA

| Model | Accuracy | Precision | Recall | F1 |
|---|---|---|---|---|
| FANCI | 0.9010 | 0.9012 | 0.9010 | 0.9010 |
| LSTM | 0.9205 | 0.9220 | 0.9205 | 0.9204 |
| Bi-LSTM | 0.9230 | 0.9232 | 0.9230 | 0.9230 |
| CNN | 0.9325 | 0.9335 | 0.9325 | 0.9325 |
| CNN-LSTM | **0.9345** | **0.9355** | **0.9345** | **0.9345** |
| LSTM-Att | 0.9080 | 0.9083 | 0.9080 | 0.9080 |
| GLHNN | 0.9275 | 0.9309 | 0.9275 | 0.9274 |

***Evaluating AGD Detection Algorithms Using DGAs***. Table 8 shows the AGD detection accuracy using DGAs, where the numbers following the accuracy values are the algorithm ranking in descending order of detection accuracy. If the mean ranking indicates the overall performance (in the last row of Table [detection]), the detection models in descending order of performance are GLHNN > CNN-LSTM > CNN > Bi-LSTM = LSTM-Att > LSTM > FANCI.

Table 8. AGD Detection Accuracy Using DGAs

| DGA | FANCI | LSTM | Bi-LSTM | CNN | CNN-LSTM | LSTM-Att | GLHNN |
|---|---|---|---|---|---|---|---|
| Bamital | 1.0000 | 1.0000 | 1.0000 | 1.0000 | 1.0000 | 1.0000 | 1.0000 |
| Banjori | 0.9785(7) | 0.9940(5) | 0.9960(4) | 0.9980(2) | 0.9980(2) | 0.9890(6) | **1.0000**(1) |
| Bedep | 0.9465(7) | 0.9760(6) | 0.9800(5) | 0.9885(3) | 0.9890(2) | 0.9835(4) | **0.9905**(1) |
| Conficker | 0.9105(6) | 0.9165(5) | 0.9385(4) | 0.9525(2) | 0.9510(3) | 0.8935(7) | **0.9530**(1) |
| DeepDGA | 0.9805(7) | 0.9905(6) | 0.9955(5) | **0.9985**(1) | 0.9965(3) | 0.9965(3) | **0.9985**(1) |
| Dyre | 1.0000 | 1.0000 | 1.0000 | 1.0000 | 1.0000 | 1.0000 | 1.0000 |
| HMM-based DGA | 0.9010(7) | 0.9205(4) | 0.9325(2) | 0.9080(4) | **0.9345**(1) | 0.9080(6) | 0.9275(3) |
| Matsnu | 0.8560(7) | 0.8870(4) | 0.8870(4) | 0.9035(3) | 0.9135(2) | 0.8800(6) | **0.9165**(1) |
| Pushdo | 0.8920(7) | 0.9275(5) | 0.9257(6) | 0.9775(3) | **0.9785**(1) | 0.9645(4) | 0.9780(2) |
| Ramdo | 0.9860(7) | 0.9995(4) | 0.9995(4) | 0.9995(4) | **1.0000**(1) | **1.0000**(1) | **1.0000**(1) |
| Suppobox | 0.7880(7) | 0.9405(5) | 0.9125(6) | 0.9410(4) | 0.9605(2) | 0.9485(3) | **0.9640**(1) |
| **Mean Ranking** | 6.89(7) | 4.89(6) | 4.44(4) | 2.89(3) | 1.89(2) | 4.44(4) | **1.33**(1) |

# 4. Related Work

Some studies sought to use conventional machine learning approaches to address the AGD detection problem. That is, they first extract statistical features of domain names and then use supervised machine learning algorithms to do the classification. Yadav et al. [35] measured the character-level statistical distances between labeled

AGD and non-AGD clusters to determine an AGD. Three distances were proposed for AGD detection: Kullback–Leibler, Jaccard index, and Edit distance. And the detection algorithm is simply based on a threshold. Schuppen et al. [5] proposed a set of lexical features of domain names to identify AGDs. Conventional machine learning algorithms like random forest and support vector machine were applied using these features. However, it is challenging to manually extract informative features from domain names. So conventional machine learning-based AGD detection often has suboptimal detection performance.

In addition to the statistical features of domain names, some studies based their AGD detection methods on some supplementary information. Antonakakis et al. [36] identified similar non-existent domain (NXDomain) responses generated by the same botnet (with the same DGA algorithm). The observation gave rise to a detection scheme using the domain name features and the host features extracted from NXDomain traffic. Schiavoni et al. [37] combined the linguistic features of domain names with the IP-based features and the denylist information in characterizing DGAs. Curtin et al. [38] relied on WHOIS data as informative side information in feature engineering. Deep learning models were developed to process the domain name features as well as the WHOIS features. However, the supplementary information beyond domain name strings has limited accessibility in the real world because of privacy concerns and non-trivial data gathering overheads.

Deep learning models found their applications to AGD detection in recent years. Most of these studies focused on detecting AGDs by purely inspecting domain names. Vij et al. [6] proposed to use LSTM to detect AGDs. Attardi et al. [12] proposed bidirectional LSTM models for DGA classification. Zhou et al. [7] applied CNNs to AGD detection. Vosoughi et al. [8] proposed a hybrid CNN/LSTM-based neural networks architecture that is applicable to AGD detection. Qiao et al. [9] proposed the LSTM combined with an attention mechanism to construct the DGA classifier.

## 5. Conclusion

In this paper, we proposed a new deep learning-based AGD detection model. The proposed detection model was experimentally validated using representative AGDs covering six classes of DGAs. The proposed detection model was compared with the state-of-the-art detection models and demonstrated the best overall detection performance among these tested models.